\documentclass[prl,aps,tightenlines,floatfix,showpacs,twocolumn]{revtex4}
\usepackage{graphicx}
\usepackage{bm}
\usepackage{amsmath}
\usepackage{amssymb}

\begin{document}

\title{Algebraic density functionals}

\author{B. G. Giraud}
\affiliation{Institut de Physique Th\'eorique,
Centre d'Etudes Saclay, 91190 Gif-sur-Yvette, France}
\email{bertrand.giraud@cea.fr}
\author{S. Karataglidis}
\affiliation{Department of Physics, University of Johannesburg,
P. O. Box 524, Auckland Park, 2006, South Africa}
\email{stevenka@uj.ac.za}

\begin{abstract}
A systematic strategy for the calculation of density functionals (DFs)
consists in coding informations about the density and the energy
into polynomials of the degrees of freedom of wave functions. DFs and 
Kohn-Sham potentials (KSPs) are then obtained by standard elimination
procedures of such degrees of freedom between the polynomials. Numerical
examples illustrate the formalism.
\end{abstract}
\date{\today}
\maketitle

Existence theorems \cite{HK} for DFs do not provide directly constructive
algorithms. Fortunately, the Kohn-Sham (KS) method \cite{KS} spares the
construction of a ``kinetic functional'' and reduces energy and
density calculations to the tuning of a local potential, $v_{KS}(r).$ 
Hence, a considerable amount of work has been dedicated to detailed 
estimates of electronic correlation energies and the corresponding
KSPs, see for instance \cite{HarJon,SolSav,TCH}. Many authors were also
concerned with representability and stability questions, see for instance
\cite{CCR} and, for calculations in subspaces, see \cite{Har} and \cite{PBLB}.
For cases where the mapping between potential and density shows singularities,
see \cite{KAD}. For reviews of the rich multiplicity of derivations of DFs and
KS solutions and their properties, we refer to \cite{DG} and \cite {ParYan},
and, for nuclear physics, to \cite{DFP}.

Local or quasi local approximations use the continuous infinity of values 
$\rho(\mathbf{r}), \forall  \mathbf{r}$, as the parameters of the
problem. However, whether for atoms, molecules or nuclei, a finite number
of parameters is enough to describe physical situations. For instance,
Woods-Saxon nuclear profiles notoriously make good approximations,
depending only on a handful of parameters, and it is easy to add a few
parameters describing, for example, long tails and/or moderate oscillations
of the density. (High frequency oscillations are unlikely, for they might
cost large excitation energies.) We can stress here, in particular, 
the one-dimensional nature of  the radial density functional (RDF) theory 
\cite{GirRDF}, valid for nuclei and/or atoms, isolated, described by 
rotationally-invariant Hamiltonians; the constrained density
minimization of energy \cite{LevLie} returns isotropic densities, with 
radial profiles, $\rho(r), 0 \leqslant r < \infty$. The number of parameters
to describe a nuclear density, therefore, can be restricted to maybe 
$\sim 10$ at most; situations with $\sim 20$ parameters are a luxury. For 
molecules, shapes are much more numerous, but a finite, while large number 
of parameters, truncating a list of multipoles for instance, still makes a 
reasonable frame. Practical DFs, therefore, can boil down to 
\textit{functions} of a finite number of parameters. Functional variations 
can then be replaced by simple derivatives.

This Letter shows how information about both the density and the energy 
can be recast into polynomials. This allows elimination of part of the 
parameters. Further polynomial manipulations locate energy extrema. Only 
density parameters are left. The same method gives KSPs. Finally we offer 
a discussion and conclusion.

Consider a basis of $n$ orthonormalized, single-particle states, 
$\varphi_{\alpha}(\mathbf{r} \sigma \tau)$, where spin and isospin 
labels $\sigma \tau$ will be understood. The orthonormalized 
Slater determinants $\phi_i$ made out of the $\varphi_{\alpha}$'s for $N$ 
fermions make a finite subspace, of some dimension $\mathcal{N},$ in which 
eigenstates of the physical Hamiltonian $H$ can be approximated by 
configuration mixings, $\Psi=\sum_{i=1}^\mathcal{N} (C_i+i C'_i) \phi_i$. Here 
$C_i$ and $C'_i$ are the real and imaginary parts, respectively, of the
mixing coefficients, but, in practice, with real matrix elements, 
$H_{ij}=\langle \phi_i | H | \phi_j \rangle$, of the Hamiltonian $H$, 
the imaginary parts $C'_i$ vanish. Both the energy $\eta$ and
the normalization are \textit{quadratic} functions of such coefficients, 
\begin{equation}
\eta =\sum_{i,j=1}^\mathcal{N} C_i H_{ij} C_j, \text{ } \sum_{i=1}^\mathcal{N} C_i^2 = 1.
\label{qudra1and2} 
\end{equation}

Let $a_{\mathbf{r}}^{\dagger}$ and $a_{\mathbf{r}}$ be the usual creation and 
annihilation operators at position $\mathbf{r}$. Tabulate the matrix elements
$\left\langle \phi_j \left| a_{\mathbf{r}}^{\dagger} a_{\mathbf{r}} \right| \phi_j \right\rangle.$ 
The density corresponding to $\Psi$ is, again, \textit{quadratic} with respect
to the $C_i$'s,
\begin{equation}
\rho(\mathbf{r}) = \sum_{ij} C_i \left\langle \phi_j \left| a_{\mathbf{r}}^{\dagger} 
a_{\mathbf{r}} \right| \phi_j \right\rangle C_j,
\end{equation}
and any parameter that is linear with respect to moments of the
density is also a quadratic function of the $C_i$'s.

Let $\{S_{\nu}(\mathbf{r})\},\ \nu=1,\dots,\infty,$ be a complete
orthonormal set of ``vanishing average'' functions. Namely, the two sets of
conditions, $\int d \mathbf{r}\, S_{\nu}(\mathbf{r})=0, \forall \nu$,
and, $\int d \mathbf{r} S_{\mu}(\mathbf{r}) S_{\nu}(\mathbf{r}) =
\delta_{\mu \nu}, \forall \mu \nu$, are satisfied. Such sets are easy to
find; in the case of one-dimensional problems, including radial ones, they
can be implemented by means of orthogonal polynomials \cite{Gi05,Nor}
and a generalization to more dimensions is easy. Then subtract from 
$\rho$ some reference density, $\rho_0,$ obtained by some approximation
relevant for the $N$ fermions. The difference, $\Delta \rho=\rho-\rho_0$, is
of a vanishing average, since, by definition, both $\rho$ and $\rho_0$
integrate out to $N$. Then the Fourier coefficients, 
\begin{equation}
\Delta_{\nu}=\int d \mathbf{r} \; S_{\nu}(\mathbf{r}) \Delta 
\rho(\mathbf{r}),
\label{dnscnstr}
\end{equation}
define $\rho$, as $\rho =\rho_0+\sum_{n=1}^{\infty} \Delta_{\nu} S_{\nu}$.
As already stated, this expansion of $\rho$ can be truncated.
at some realistic order $\mathcal{N}'$, lower than the number of 
independent parameters $C_i$. The $\Delta_{\nu}$'s are \textit{quadratic} in the 
$C_i$'s,
\begin{equation}
\Delta_{\nu}=\sum_{ij} C_i \left[\int d \mathbf{r}\; S_{\nu}(\mathbf{r})
\left\langle \phi_j \left| a_{\mathbf{r}}^{\dagger} a_{\mathbf{r}} \right| \phi_j \right\rangle 
\right] C_j - \rho_{0 \nu} .
\label{qudra3}
\end{equation}
Note the auxiliary numbers,
$\rho_{0\nu}=\int d \mathbf{r}\, S_{\nu}(\mathbf{r})\, \rho_0(\mathbf{r})$.

It is then trivial to use the $\mathcal{N}'$ density constraints, 
Eqs. (\ref{qudra3}), and the normalization in Eqs.~(\ref{qudra1and2}), to
eliminate, for instance, the last $(\mathcal{N}'+1)$ coefficients $C_i$. This 
leaves a polynomial relation, 
$\mathcal{R}
(\eta,\Delta_1,\dots,\Delta_{\mathcal{N}'},C_1,\dots,C_{\mathcal{N}-\mathcal{N}'-1})=0$,
between the energy, the density parameters, and the remaining coefficients 
$C_i$. Finally, the energy must be minimized with respect to such remaining 
coefficients, via still polynomial conditions, 
$\partial \mathcal{R}/\partial C_i=0,\text{ } i=1,\dots,\mathcal{N}-\mathcal{N}'-1.$
This gives a polynomial relation, 
$\mathcal{E}(\eta,\Delta_1,\dots,\Delta_{\mathcal{N}'})=0$, between the energy and 
the density parameters. This polynomial $\mathcal{E}$ is our ``algebraic'' DF.
It accounts for all contributions to the energy, both without and with 
correlations, for only matrix elements of the full $H$ are used.

The procedure can be further simplified in the following way. Let $\mathcal{H}$
be the matrix representing the Hamiltonian on an orthonormal basis for a 
suitable subspace of wave functions, and, similarly, let, for instance,
$\mathcal{D}_1, \mathcal{D}_2$ be the matrices representing two constraints
selected to parametrize the density, such as, for instance, two among the
parameters $\left(\Delta_{\nu}+\rho_{0 \nu}\right)$. Set the equation,
polynomial in all three variables $\varepsilon,\lambda_1,\lambda_2$,
\begin{equation}
P\left( \varepsilon,\lambda_1,\lambda_2 \right) \equiv \det \left(\mathcal{H} - 
\lambda_1 \mathcal{D}_1 - \lambda_2 \mathcal{D}_2 - \varepsilon \right) = 0.
\label{detequat}
\end{equation} 
Here $\varepsilon$ is the free energy, lowest eigenvalue of 
$\left( \mathcal{H} - \lambda_1 \mathcal{D}_1 - \lambda_2 \mathcal{D}_2 \right)$, and 
the $\lambda$'s are Lagrange multipliers. It is well known that 
$\partial \varepsilon/\partial \lambda_i = - D_i,\, i=1,2,$ where 
$D_i \equiv \left\langle \mathcal{D}_i \right\rangle$ is the expectation value of the
corresponding constraint. From Eq. (\ref{detequat}) such partial derivatives
read, 
$\partial \varepsilon/\partial \lambda_i = -
(\partial P/\partial \lambda_i) /
(\partial P/\partial \varepsilon), \; i=1,2,$ hence two more polynomial
relations are obtained,
\begin{equation}
Q_i(D_i,\varepsilon,\lambda_1,\lambda_2) \equiv 
(\partial P/\partial \varepsilon)\, D_i - (\partial P/\partial \lambda_i)=0.
\label{partderiv}
\end{equation}
Replace in Eqs. (\ref{detequat},\ref{partderiv}) the free energy 
by its value, $\varepsilon = \eta - \lambda_1 D_1 - \lambda_2 D_2$,
in terms of the energy, $\eta \equiv \langle \mathcal{H} \rangle$ and the 
constraints, $D_1,D_2$. This creates three polynomials in terms
of $\eta,D_1,D_2,\lambda_1,\lambda_2$, out of which 
$\lambda_1,\lambda_2$ can be eliminated, for a final polynomial equation, 
$\mathcal{E}(\eta,D_1,D_2)=0$. This easy Legendre transform generates our 
``algebraic DF''. A generalization to any number of quadratic constraints is
trivial. Such algebraic DFs are not open formulae of the form,
$\eta=F(D_1,\dots,D_{\mathcal{N}'})$, but they provide roots for $\eta$ at any
realistic degree of numerical accuracy. Incidentally, they may also give
excited energies and/or spurious ones, a well known property \cite{PerLev}
of DFs.

For an illustrative toy model, we consider two fermions only and set the 
one-body part of $H$ as, $K=-d^2/(2 dr_1^2)-d^2/(2 dr_2^2)+(r_1^2+r_2^2)/2$, 
the sum of two harmonic oscillators, and its two-body part as a translation 
invariant, separable potential, defined in coordinate representation by,
\begin{multline}
\left\langle r_1 r_2 \left|V \right| r'_1 r'_2 \right\rangle = - V_0
\delta \left[ (r_1+r_2-r'_1-r'_2)/2 \right] \\
\times e^{-[(r_2-r_1)^2+(r'_2-r'_1)^2]/4} (r_2-r_1)(r'_2-r'_1)/\sqrt{2 \pi}.
\label{separpot}
\end{multline} 
Then, given the first 4 wave functions, $\varphi_0,\dots,\varphi_3$, of the 
one-dimensional harmonic oscillator, we create, to prepare a configuration 
mixing, a basis of 4 negative parity Slater determinants. These read, in a 
transparent notation, $\{ \varphi_0, \varphi_1\}, \{\varphi_0, \varphi_3\}, 
\{\varphi_2, \varphi_1\}, \{\varphi_2, \varphi_3\}$. We set $V_0=3$ for a
numerical test. To constrain $H$, we choose the second moment operator, 
$r_1^2+r_2^2$. The matrices representing $H$ and the constraint in the toy
subspace read, 
\begin{equation}
\mathcal{H}=\left[ 
\begin{matrix}
 -1 & 0            & 0            & 0   \cr 
  0 & 7/4          & 3 \sqrt{3}/4 & 0   \cr 
  0 & 3 \sqrt{3}/4 & 13/4         & 0   \cr 
  0 & 0            & 0            & 45/8
\end{matrix}
         \right],
\end{equation}
and
\begin{equation} 
\mathcal{D}=\left[ \begin{matrix}
 2          & \sqrt{3/2} & \sqrt{1/2} & 0          \cr
 \sqrt{3/2} & 4          & 0          & \sqrt{1/2} \cr
 \sqrt{1/2} & 0          & 4          & \sqrt{3/2} \cr
 0          & \sqrt{1/2} & \sqrt{3/2} & 6          
\end{matrix} \right].
\end{equation}
The equations which correspond to Eqs.~(\ref{detequat}),(\ref{partderiv}) read,
\begin{multline}
P_{\text{toy}}(\varepsilon,\lambda)=
-360 + 154 \varepsilon + 344 \varepsilon^2 - 154 \varepsilon^3 + 
 16 \varepsilon^4 + 1464 \lambda \\ + 1692 \varepsilon \lambda - 
 1636 \varepsilon^2 \lambda + 256 \varepsilon^3 \lambda + 725 \lambda^2 - 
 5140 \varepsilon \lambda^2 + \\
 1408 \varepsilon^2 \lambda^2 - 4192 \lambda^3 + 
 3072 \varepsilon \lambda^3 + 2064 \lambda^4 =0, \\
Q_{\text{toy}}(D,\varepsilon,\lambda)=
-1464 - 1692 \varepsilon + 1636 \varepsilon^2 - 256 \varepsilon^3 -
 1450 \lambda + \\ 10280 \varepsilon \lambda - 2816 \varepsilon^2 \lambda +
 12576 \lambda^2 - 9216 \varepsilon \lambda^2 - 8256 \lambda^3 + \\
(154 +  688 \varepsilon - 462 \varepsilon^2 + 64 \varepsilon^3 + 
 1692 \lambda - 3272 \varepsilon \lambda + \\ 768 \varepsilon^2 \lambda - 
 5140 \lambda^2 + 2816 \varepsilon \lambda^2 + 3072 \lambda^3)D=0.
\end{multline}
Finally, the substitution, $\varepsilon=\eta-\lambda  D$, followed by the 
elimination of $\lambda$, generates the desired polynomial equation, 
$\mathcal{E}_{\text{toy}}(\eta,D)=0$. (This polynomial $\mathcal{E}_{\text{toy}}$ is of
order $12$ in both $\eta$ and $D$ and is a little cumbersome for a publication
here. It is available to the interested reader.)

\begin{figure}
\scalebox{0.65}{\includegraphics*{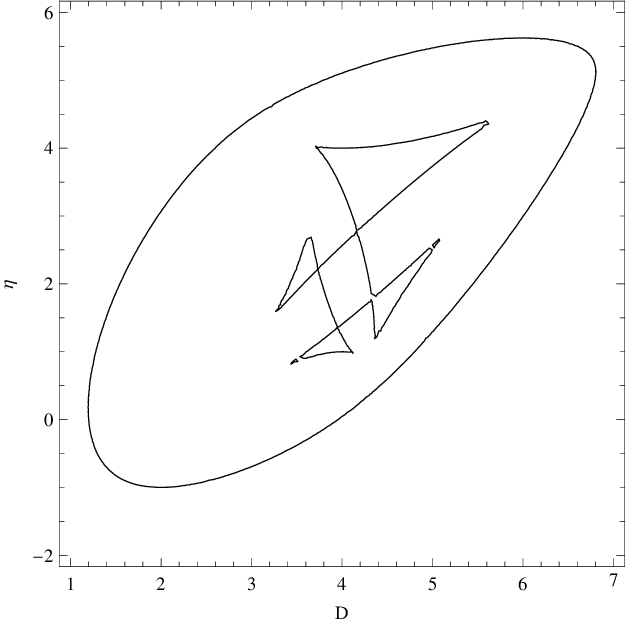}}
\caption{\label{figure1}Contour $\mathcal{E}_{\text{toy}}(\eta,D)=0$ for the
configuration mixing model with $4 \times 4$ matrices, as described in the
text.}
\end{figure}

We show in Fig. \ref{figure1} the contour line, $\mathcal{E}_{\text{toy}}(\eta,D)=0.$ The ground
state is found at the lowest point of the oval envelope, with coordinates,
$D=2, \eta=-1$. The highest and lowest eigenvalues of $\mathcal{H}$ are, $45/8$
and  $-1$, and those of $\mathcal{D}$ are, $4 \pm \sqrt{4+\sqrt{15}}$, namely
$\sim 6.81$ and $\sim 1.19$. This is confirmed by the extremal points, up,
down, right and left, of the oval. The inside pattern refers to excited
states. The concavity of the lowest part of the envelope and convexity of
its highest part are transparent properties of the theory. They generalize
for any dimension of the subspace and any number of constraints; we tested
this generalization with further toy models. Moreover, when, via 
embedded subspaces, the dimension $\mathcal{N}$ of the matrices, 
$\mathcal{H}, \mathcal{D}_i$, grows while $H$ and the constraints are kept the
same, a growth of the envelope is found and the bottom of the envelope
converges towards a limit, as expected. This gives numerical estimates for
an extrapolation of this concave part towards its limit for 
$\mathcal{N} \rightarrow \infty$.

Such concavities should also occur in DF theories with a continuous
infinity of constraints. But they are often difficult to verify, and are,
therefore, overlooked, although they are an important test of soundness.

A byproduct of the procedure consists of a polynomial relating the
potential energy to the constraints. Set the Hamiltonian as, 
$H=h+V$, with $V=-V_0 \mathcal{V}$, where $V_0$ is an interaction strength
and $\mathcal{V}$ gives all details of interaction shapes. Nothing prevents one
from considering $V_0$ as a Lagrange multiplier and obtain, via the polynomial
method pushed one step further, a polynomial, 
$\mathcal{F}
(\langle h \rangle,\langle \mathcal{V} \rangle,D_1,\dots,D_{\mathcal{N}'})$, linking
$\langle h \rangle$ to the expectation values of $\mathcal{V}$ and
the constraints. A standard result of this Legendre transform is, 
$\partial \langle h \rangle / \partial \langle \mathcal{V} \rangle = V_0$,
\textit{i.e.},
\begin{multline}
\mathcal{G}
(V_0,\langle h \rangle,\langle \mathcal{V} \rangle,D_1,\dots,D_{\mathcal{N}'})
\equiv \\  (\partial \mathcal{F}/\partial \langle h \rangle)\, V_0 -
\partial \mathcal{F} / \partial \langle \mathcal{V} \rangle =0.
\label{poteq}
\end{multline}
Replace, in $\mathcal{F}$ and $\mathcal{G}$, the quantity $\langle h \rangle$
by $\eta+\langle \mathcal{V} \rangle V_0.$ Then eliminate $\eta$ and $V_0$
between $\mathcal{E}$ and such modified $\mathcal{F}$ and $\mathcal{G}$. This links
$\langle \mathcal{V} \rangle$, hence $\langle V \rangle$, to the $D_i$.
It must be stressed here that now $\langle \mathcal{V} \rangle$ should \textit{not}
be  minimized with respect to the $D_i$; rather, those $D_i$ values to be 
used are those that minimize the total energy $\eta$.

A similar argument provides the kinetic energy, or any other part of $\eta$,
in the same context of total energy constrained minimization. Such results
are of interest for a detailed analysis of corrections induced by
correlations.

The direct approach resulting from Eqs. (\ref{detequat}) and (\ref{partderiv})
bypasses the KS approach. For the sake of completeness, however, we now 
show how this theory can handle determinants and also calculate a KSP. 
Consider a basis of $n$ single-particle states, 
$\varphi_{\alpha},\, \alpha=1,\dots,n$, a Slater determinant $\Phi$ made
of $N$ orthonormal orbitals,
$\psi_{\gamma}=\sum_{\alpha=1}^n c_{\gamma \alpha} \varphi_{\alpha}$, and a 
Hamiltonian with its one-body and two-body parts, $H=K+V$, assuming real
matrix elements,
$K_{\alpha \beta}=\langle \varphi_{\alpha} | K | \varphi_{\beta} \rangle$ and 
$V_{\alpha \beta \gamma \delta}=\langle \varphi_{\alpha} \varphi_{\beta} | V |
\varphi_{\gamma} \varphi_{\delta} \rangle.$ The energy of $\Phi$ becomes
quartic in  the orbital coefficients, $c_{\gamma \alpha}$, because of $V$,
and even needs order 6 if three-body forces are introduced, but the $N(N+1)/2$
orthonormalization constraints and the density remain quadratic. Obviously,
a few parameters constraining the density of $\Phi$, or its difference from
some $\rho_0,$ can again be chosen as quadratic in the coefficients 
$c_{\gamma \alpha}$. Elementary eliminations then yield a polynomial
relation between Slater energy and density parameters.

The following toy model, in which the number of free parameters reduces to
$\mathcal{N}=2$ and we choose that of density constraints as $\mathcal{N}'=1$,
illustrates the strategy. From the first 4 wave functions,
$\varphi_0,\dots,\varphi_3$, of the one-dimensional harmonic oscillator,
set a Slater determinant $\Phi$ made of one positive and one negative parity
orbitals,
\begin{equation}
\psi_+ = t\varphi_0 + u\varphi_2, \text{ }  
\psi_- = v \varphi_1 + w \varphi_3.
\label{orbits}
\end{equation}
One constraint is spared by such orbital parities, which ensure orthogonality.
Normalization constraints can also be spared if they are ensured by a 
``trigonometric'' form of the components, 
$t=(1-a^2)/(1+a^2), u=2a/(1+a^2), v=(1-b^2)/(1+b^2), w=2 b/(1+b^2)$,
with both parameters, $a,b$, real numbers. The density is a Gaussian
modulated by a polynomial,
\begin{equation}
\rho(r)=\pi^{-\frac{1}{2}} e^{-r^2} \left( a_6 r^6+a_4 r^4+a_2 r^2+a_0 \right),
\label{dnstpoly}
\end{equation}
with two independent coefficients only, because of the two parameters only,
$a,b$, for $\Phi$. One of the relations between $a_6,\dots,a_0$ is linear, 
since the integral,
\begin{equation}
\int_{-\infty}^{\infty} dr\, \rho(r)=\frac{15}{8} a_6+\frac{3}{4} a_4+
\frac{1}{2} a_2 + a_0 = 2,
\label{massconsrv}
\end{equation}
must equate to the particle number. The other comes from the condition that
gives the density of $\Phi$,
\begin{equation}
|\psi_+|^2 + |\psi_-|^2 = \pi^{-\frac{1}{2}}
e^{-r^2} (a_6\, r^6 + a_4\, r^4 + a_2\, r^2+a_0).
\label{fitrho}
\end{equation}
Insert Eqs.~(\ref{orbits}) into Eq.~(\ref{fitrho}) and take advantage of
the harmonic oscillator basis states. The density constraint, 
${\Phi \Rightarrow \rho}$, then means 4 conditions in terms of $t,u,v,w$,
\begin{align}
4\, w^2/3 = a_6,\ \ \ \ t^2 - \sqrt{2}\, t\, u + u^2/2 = a_0, \nonumber \\
2\, u^2 + 4\, \sqrt{2/3}\, v\, w - 4\, w^2 = a_4,  \nonumber \\
2\, \sqrt{2}\, t\, u - 2\, u^2 + 2\, v^2 - 2\, \sqrt{6}\, v\, w + 3\, w^2 = 
a_2.
\label{fita0246}
\end{align}
We can use these, Eqs. (\ref{fita0246}), rather then Eqs. (\ref{qudra3}), for 
our argument. In terms of $a,b,$ these Eqs. (\ref{fita0246}) read,
\begin{multline}
a_6  = \frac{16\, b^2}{3\, (1+b^2)^2},\text{ }
a_0 = \frac{1 - 2\, \sqrt{2} a + 2\, \sqrt{2} a^3 + a^4}{(1+a^2)^2}, \\
3\ a_4\ (1+a^2)^2\ (1+b^2)^2 = 8\, (
 3\, a^2 + \sqrt{6} b\ + \\ 2 \sqrt{6}\, a^2 b + \sqrt{6}\, a^4 b -
 6\, b^2 - 6\, a^2 b^2 - 6\, a^4 b^2 - \\ \sqrt{6}\, b^3 -
 2 \sqrt{6}\, a^2 b^3 - \sqrt{6}\, a^4 b^3 + 3\, a^2 b^4), \\
a_2\, (1+a^2)^2\, (1+b^2)^2 = 2\, (
1 + 2 \sqrt{2}\, a - 2\, a^2 - 2 \sqrt{2}\, a^3 + \\ a^4 - 2 \sqrt{6} b -
 4 \sqrt{6}\, a^2 b - 2 \sqrt{6}\, a^4 b + 4\, b^2 + 4 \sqrt{2}\, a b^2 - \\
 4 \sqrt{2}\, a^3 b^2 + 4\, a^4 b^2 + 2 \sqrt{6}\, b^3 + 4 \sqrt{6}\, a^2 b^3
 + 2 \sqrt{6}\, a^4 b^3 +  \\ b^4 + 2 \sqrt{2}\, a b^4 - 2\, a^2 b^4 - 
 2 \sqrt{2}\, a^3 b^4 + a^4 b^4).
\label{a6042ab}
\end{multline}
For the sake of simplicity, we select $a_6$ and $a_0$ as primary, independent
parameters of $\rho$ and eliminate $a,b$ between those of Eqs. (\ref{a6042ab})
that give $a_6,a_0,a_4$. The result,
\begin{multline}
 256 - 1024 a_0 + 1536 a_0^2 - 1024 a_0^3 + 256 a_0^4 - 768 a_4 + \\
 1792 a_0 a_4 - 1280 a_0^2 a_4 + 256 a_0^3 a_4 + 864 a_4^2 - 960 a_0 a_4^2 +
 \\ 352 a_0^2 a_4^2 -  432 a_4^3 + 144 a_0 a_4^3 + 81 a_4^4 - 4608 a_6 + \\
 3840 a_0 a_6 - 2048 a_0^2 a_6 + 768 a_0^3 a_6 + 8640 a_4 a_6 - \\
 6912 a_0 a_4 a_6 + 2112 a_0^2 a_4 a_6 - 5184 a_4^2 a_6 + 
 1296 a_0 a_4^2 a_6 +  \\ 972 a_4^3 a_6 + 25056 a_6^2 - 10944 a_0 a_6^2 +
 1824 a_0^2 a_6^2 - \\ 22032 a_4 a_6^2 + 4752 a_0 a_4 a_6^2 +
 5346 a_4^2 a_6^2 - 38880 a_6^3 + \\ 6480 a_0 a_6^3 + 14580 a_4 a_6^3 +
 18225 a_6^4 = 0,
\label{a604}
\end{multline}
completes Eq. (\ref{massconsrv}) to link $a_4,a_2$ to $a_6,a_0.$ Incidentally,
Eqs. (\ref{a6042ab}) show that $0 \leq 3 a_6 \leq 4$ and $0 \leq 2 a_0 \le 3.$

The same toy Hamiltonian $H$ as was used to generate Fig. 1 induces the
Slater energy,
\begin{multline}
\eta \equiv 
\langle \Phi |H| \Phi \rangle = (t^2+5u^2+3v^2+7w^2)/2\, - V_0/8\ \times \\ 
\left[ 2\, (4\, t^2+u^2)\, v^2 - 4 \sqrt{3}\, t\, u\, v\, w\, + (6\, t^2+u^2)
\, w^2 \right] = \\
 [\, 2 + 12 a^2 + 2 a^4 + 12 b^2 + 40 a^2 b^2 + 12 a^4 b^2 + 2 b^4 + \\
 12 a^2 b^4 + 2 a^4 b^4 - V_0\, (1 -  a^2 + a^4 - 2 \sqrt{3} a b + 
 2 \sqrt{3} a^3 b + b^2 - \\ 2 a^2 b^2 + a^4 b^2 + 2 \sqrt{3} a b^3 - 
 2 \sqrt{3} a^3 b^3 + b^4 - a^2 b^4 + a^4 b^4) \, ]\, / \\
[(1+a^2)\, (1+b^2)]^2.
\label{etaab}
\end{multline}

Given $H,$ the DF is defined by the constrained minimization \cite{LevLie},
$\mathcal{F}[\rho] =  \mathrm{Min}_{\Phi \Rightarrow \rho}\, 
\left\langle \Phi \left| H \right| \Phi \right\rangle,$
where the constraint, ${\Phi \Rightarrow \rho},$ will now be interpreted as
just a constraint ${\Phi \Rightarrow a_6}.$ We motivate this choice of the
maximum degree coefficient by at least two reasons, namely, i) it is an 
interesting degree of freedom, since it can be interpreted as a 
``halo driving'' parameter, ii) it will actually turn out that the 
ground state corresponds to $a_6=0,$ (hence, no halo!), this value $0$
interestingly sitting on an edge of the convex domain of densities;
variational calculus at edges of domains are notoriously challenging.
We can, therefore, eliminate $b$ between Eq. (\ref{etaab}) and the first among
Eqs. (\ref{a6042ab}). This implements the constraint, 
${\Phi \Rightarrow a_6},$ in a precursor situation before energy minimization 
with respect this constrained $\Phi,$ whose last free parameter is $a.$ This 
``precursor'' energy is given by,
\begin{multline}
\mathcal{P}(\eta,a_6,a)= 1024 + 12288 a^2 + 38912 a^4 + \\ 
 12288 a^6 + 1024 a^8 + 1536 a_6 + 12288 a^2 a_6 + \\ 
 21504 a^4 a_6 + 12288 a^6 a_6 + 1536 a^8 a_6 + 576 a_6^2 + \\
 2304 a^2 a_6^2 + 3456 a^4 a_6^2 + 2304 a^6 a_6^2 + 576 a^8 a_6^2 - \\
 1024 \eta - 8192 a^2 \eta - 14336 a^4 \eta - 8192 a^6 \eta - \\
 1024 a^8 \eta - 768 a_6 \eta  - 3072 a^2 a_6 \eta - 4608 a^4 a_6 \eta - \\
 3072 a^6 a_6 \eta - 768 a^8 a_6 \eta + 256 \eta^2 + 1024 a^2 \eta^2 + \\ 
 1536 a^4 \eta^2 +  1024 a^6 \eta^2 + 256 a^8 \eta^2 - 1024 V_0 - \\
 5120 a^2 V_0 + 4096 a^4 V_0 - 5120 a^6 V_0 - 1024 a^8 V_0 - \\
 576 a_6 V_0 + 384 a^2 a_6 V_0 + 384 a^4 a_6 V_0 + 384 a^6 a_6 V_0 - \\
 576 a^8 a_6 V_0 + 144 a_6^2 V_0 + 288 a^2 a_6^2 V_0 + 288 a^4 a_6^2 V_0 + \\
 288 a^6 a_6^2 V_0 + 144 a^8 a_6^2 V_0 + 512 \eta V_0 + 512 a^2 \eta V_0 + \\
 512 a^6 \eta V_0 + 512 a^8 \eta V_0 - 96 a_6 \eta V_0 - 192 a^2 a_6 \eta V_0
 - \\
 192 a^4 a_6 \eta V_0 - 192 a^6 a_6 \eta V_0 - 96 a^8 a_6 \eta V_0 + 256 V_0^2
 - \\
 512 a^2 V_0^2 + 768 a^4 V_0^2 - 512 a^6 V_0^2 + 256 a^8 V_0^2 - \\
 96 a_6 V_0^2 - 480 a^2 a_6 V_0^2 + 960 a^4 a_6 V_0^2 - 480 a^6 a_6 V_0^2 - \\
 96 a^8 a_6 V_0^2 + 9 a_6^2 V_0^2 + 432 a^2 a_6^2 V_0^2 - 846 a^4 a_6^2 V_0^2
 + \\
 432 a^6 a_6^2 V_0^2 + 9 a^8 a_6^2 V_0^2 = 0.
\label{precurso}
\end{multline}
This is now combined with the energy minimization,
$\partial \mathcal{P}/\partial a=0,$
with respect to $a,$ thus eliminating $a,$
\begin{multline}
\mathcal{Q}(\eta,a_6)=(32 + 24 a_6 - 16 \eta - 16 V_0 + 3 a_6 V_0)\ \times \\
 (128 + 48 a_6 - 32 \eta - 8 V_0 + 3 a_6 V_0)\ (4096 + \\ 4608 a_6 + 
 1152 a_6^2 - 3072 \eta - 1536 a_6 \eta + 512 \eta^2 - \\ 2304 V_0 -
 480 a_6 V_0 + 216 a_6^2 V_0 + 640 \eta V_0 - \\ 144 a_6 \eta V_0 + 
  128 V_0^2 -  144 a_6 V_0^2 + 63 a_6^2 V_0^2) = 0.
\label{DF6}
\end{multline}
This polynomial $\mathcal{Q}$, Eq. (\ref{DF6}), is an algebraic DF for the 
Slater $\Phi$. In turn, with a final minimization, $\partial Q/\partial a_6=0$,
``with respect to the density'', actually here w.r.t. the $a_6$ parameter,
the polynomial equation for $\eta$ reads,
\begin{multline}
\mathcal{S}(\eta)=(\eta + V_0 - 2)\, (4 \eta + V_0 - 16)\, (8 \eta + V_0 - \\ 48)
\, 
 (4 \eta + 3 V_0 - 16)\, (64 + 36 V_0 - 2 \eta V_0 + V_0^2) \times \\
 (-1024 - 1152 V_0 + 64 \eta V_0 + 316 V_0^2 - 348 \eta V_0^2 + \\
 47 \eta^2 V_0^2 - 264 V_0^3 + 52 \eta V_0^3 + 5 V_0^4) = 0.
\label{solventeta}
\end{multline}
An elimination of $\eta$ between the same conditions, $\mathcal{Q}=0$ and 
$\partial \mathcal{Q}/\partial a_6=0,$ yields the condition for $a_6,$
\begin{multline}
a_6\ (3 a_6 -4)\ (3 a_6 V_0 - 24 V_0 -64)\ (524288 + \\ 491520 V_0 -
 49152 a_6 V_0 + 151552 V_0^2 + 46080 a_6 V_0^2 - \\ 54144 a_6^2 V_0^2 + 
 18432 V_0^3  +  7680 a_6 V_0^3 - 10152 a_6^2 V_0^3 + \\
 3024 a_6 V_0^4 - 2961 a_6^2 V_0^4)=0.
\label{solventa_6}
\end{multline}

For the numerical illustrations that follow, set $V_0=3.$ Then
Eq. (\ref{DF6}) becomes, 
\begin{multline}
(16 - 33 a_6 + 16 \eta)\, (104 + 57 a_6 - 32 \eta)\, (-1664 + 1872 a_6 + \\
   2367 a_6^2 - 1152 \eta - 1968 a_6 \eta + 512 \eta^2)=0.
\label{DF63}
\end{multline}
For $V_0=3$ the lowest root of Eq. (\ref{solventeta}) is, $\eta=-2.98623.$ But
it is soon recognized as spurious, because, inserted into Eq. (\ref{DF6}), 
it returns absurd, negative only values of $a_6.$ This is confirmed by a 
detailed consideration, in the only allowed domain, $0 \le a_6 \le 4/3,$ of
the solution branches yielded by Eq. (\ref{DF63}), namely 
$\eta=-1+33 a_6/16,$ $\eta=13/4+57 a_6/12,$ 
$\eta=(72 + 123 a_6 \pm \sqrt{18496 + 2736 a_6 - 3807 a_6^2})/64.$ These are
shown in Fig. 2, and clearly validate the second lowest root, $\eta=-1,$ of 
Eq. (\ref{solventeta}), together with that root, $a_6=0,$ of 
Eq. (\ref{solventa_6}), hence $b=0.$
 
It is then trivial to take advantage of Eq. (\ref{etaab}) by inserting the
results, $\eta=-1,$ $b=0,$ and obtain, $a=0,$ hence $a_0=1,$ then $a_4=0$ and
$a_2=2.$ The optimal density is, therefore, 
$\rho=(1+2 r^2)\, e^{-r^2}/\sqrt{\pi}$. Notice, incidentally, that we have five
equations at our disposal, namely Eq. (\ref{etaab}) and Eqs. (\ref{a6042ab})
to directly relate $\eta$ and the $a_i$'s via an elimination of $a,b,$ 
via polynomial conditions of the form $\mathcal{T}(\eta,a_6,a_4)=0,$ and
$\mathcal{U}(\eta,a_6,a_2)=0,$ for instance. We verified that the same set,
$\{a_0=1,a_2=2,a_4=0,a_6=0\},$ results from such  a direct use of the values,
$V_0=3,\, \eta=-1.$ 

\begin{figure}
\scalebox{0.70}{\includegraphics*{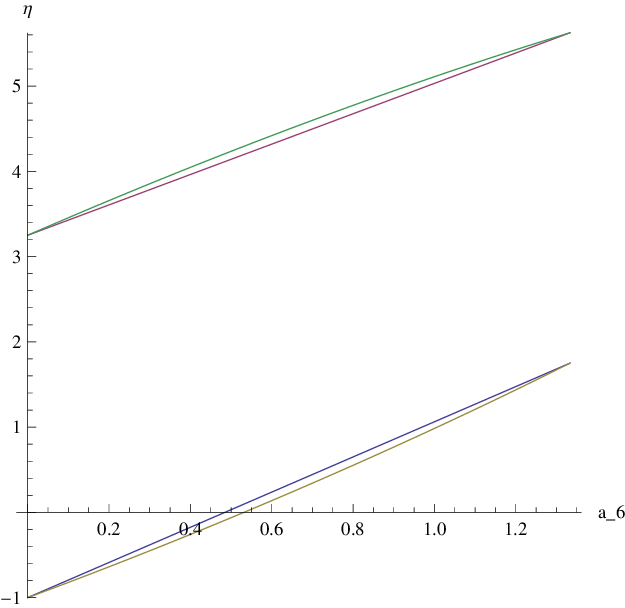}}
\caption{\label{figure2}Trajectories $\eta(a_6),$ see Eq. (\ref{DF6}), for
the toy model for a Slater determinant, as described in the text.}
\end{figure}

While usually many wave functions can give the same density, this toy model
allows the wave function to be identified. It is that Slater determinant 
$\Phi_{gs}$ made of $\varphi_0$ and $\varphi_1.$ This $\Phi_{gs}$ is,
obviously, the first two-fermion eigenstate of our $K,$ a sum of two harmonic
oscillators, with eigenvalue, $2=1/2+3/2.$ The same $\Phi_{gs}$ is also an
eigenstate of $V,$ Eq. (\ref{separpot}), since in the following Jacobi
coordinates, $R=(r_1+r_2)/2,$ $r_{21}=r_2-r_1,$ the wave function of 
$\Phi_{gs}$ reads, $\propto  \exp(-R^2)\ r_{21}\, \exp(-r_{21}^2/4),$ while
the representation of $V$ is,
$\langle R r_{21} |V| R' r'_{21} \rangle \propto \delta(R-R')\ 
r_{21}\, r'_{21}\, \exp[-(r_{21}^2+r_{21}^{\prime 2})/4],$ showing an obvious 
projector on that relative motion expressed by $\Phi_{gs}.$ The corresponding 
eigenvalue is, $-V_0,$ hence our result, $\eta=-1,$ when $V_0=3.$ We took
great care to verify that the same results are obtained if, instead of $a_6,$
we use other choices for ``density constraint'', such as the parameter $a_0,$
or a moment such as the second one, $a_0/2 + 3 a_2/4 + 15 a_4/8 + 105 a_6/16,$
or the local value $\rho(\theta)$ at some testing point $r=\theta.$ Such
rearrangements of information with respect to the wave function
parameters may be of some interest for questions of physics or numerical 
convenience, but do not change the nature of the algebra nor the the final
results. It can be noted here that what is important for the method is that
the energy and the constraints be polynomials of the parameters. The fact 
that, in the toy model, the density is described by a polynomial of $r$ is
not essential. It only makes the algebra slightly simpler. With wave functions
more complicated than harmonic oscillator ones, any choice of moments, or 
local values of $\rho,$ still makes eligible constraints.

An issue which will arise in all future models using this polynomial method is
that the final minimization of $\eta$ must be performed within a convex domain
of densities: what conditions must the coefficients $a_i,$ or those other
selected parameters (moments, local values, etc), satisfy to maintain 
$\rho$ positive? This question was recently \cite{GiPe} solved by means of the
Sturm criterion, for a general class of positive functions having positive
Fourier transforms. The criterion gives the number of real roots of a
polynomial, and can be used to ensure that a polynomial has no real roots.
As seen in the toy model, the detailed structure of the calculation
can be a guide to define the physically acceptable domain of parameters,
see the bounds found for $a_6$ and $a_0.$ For more subtle questions about the
topology of acceptable functional spaces of densities and trial functions,
we refer to \cite{UllKohtopo}, but will conjecture, without proof, that here
with traditional functions (harmonic, Coulomb) and their configuration
mixings, the positivity of $\rho$ should be sufficient.

There is also the question of spurious solutions. The elimination of that
spurious solution found in the Slater toy model turned out to be trivial. For
more complicated systems, spurious solutions \cite{Har,PBLB} might certainly
pop up, but an analysis for their detection remains easy.  In particular, for
other toy models that we tested, spurious solutions were found to induce
values of physical parameters out of their allowed range, and/or even complex
values while only real ones are acceptable. We can insist that the final,
polynomial equation for the energy, $\mathcal{S}(\eta)=0,$  can only create a
finite number of candidate solution branches to be investigated.

This concludes our toy model as a demonstration of a handling of determinants 
in this algebraic approach. But we can still take advantage of it
for a study of the ``kinetic Kohn-Sham functional''. First notice that the
``harmonic energy'', $\langle \Phi | K | \Phi \rangle,$ and the kinetic energy
of $\Phi$ differ by only an explicit functional of the density, namely half
of its second moment, $\int dr\, r^2\, \rho(r).$ The search for a functional 
for $\langle \Phi | K | \Phi \rangle,$ therefore, is a problem equivalent to 
that for the kinetic energy. Set now $V_0=0$ in Eqs. (\ref{etaab}). 
The same program of elimination that was used for a full energy functional
now returns a simpler, and very transparent, form of Eq. (\ref{solventeta}),
$(\eta-6) (\eta-4)^2 (\eta-2)=0.$ The corresponding version of
Eq. (\ref{DF6}), $(2 \eta - 4 - 3 a_6)^2\, (2 \eta - 8 - 3 a_6)^2=0,$ gives
$\eta=2$ and $\eta=4$ if $a_6=0.$ This means determinants made of
$\{\varphi_0,\varphi_1\}$ and $\{\varphi_2,\varphi_1\},$ respectively. For
$a_6=4/3,$ at the other edge of the domain, the harmonic energies are $\eta=4$
and $\eta=6,$ with determinants $\{\varphi_0,\varphi_3\}$ and 
$\{\varphi_2,\varphi_3\},$ respectively.

After this proof that the method is basically the same for determinants as 
for configuration mixings, we can stress that configuration mixings have
the technical advantage that the energy is quadratic only and permits the
short cut described at the stage of Eqs (\ref{detequat},\ref{partderiv}).

A constructive derivation of KSPs is available. For instance, truncate some 
single particle basis and let $\mathcal{P}$ be the projector upon the 
resulting, finite dimensional subspace for a system of $N$ fermions, with 
their Hamiltonian $H$, or rather now, $\mathcal{P} H \mathcal{P}$. Given the 
kinetic energy operator $T,$  choose a local potential $w_0(r),$ hence a 
one-body operator $W_0=\sum_{i=1}^N w_0(r_i)$, hence a one-body Hamiltonian 
$H_0=T+W_0$, so that the ground state of 
$\mathcal{P} H_0 \mathcal{P}$, a Slater determinant $\Phi_0$, be non 
degenerate and providing an approximate density $\rho_0$ for the system. 
For any density $\rho$ in the subspace, the integral, $\int \Delta \rho$, 
of the difference, $\Delta \rho=\rho-\rho_0$, vanishes as already stated.
(Here and in the following, the integral sign, $\int$, means 
$\int r^{d-2}\, dr$ depending on the $d$-dimensional problem under 
consideration.)  Expand, as already discussed, $\Delta \rho$
in a basis of orthonormal functions $S_{\beta}(r)$, ``constrained by
vanishing  averages'' \cite{Gi05,Nor},
$\Delta \rho(r)=\sum_{\beta=1}^{\infty} b_{\beta}\, S_{\beta}(r).$
Truncate the expansion at some suitable order $\mathcal{N}'.$
Again, given a determinant $\Phi$ with the parameters $c_{n \ell m}^{\alpha}$
of its orbitals, or given a correlated state,
$\Psi=\sum_q C_q\, \Phi_q$, the constraints, $\Phi \Rightarrow b_{\beta}$ or
$\Psi \Rightarrow b_{\beta}$, are polynomials of the parameters. Given
$H_0$, the polynomial method returns a polynomial
$\mathcal{K}(\kappa,b_1,\dots,b_{\mathcal{N}'})$
for a reference functional, such that the lowest root of the equation, 
$\mathcal{K}=0$, represents the constrained minimum,
$\kappa' = \mathrm{Min}_{\Phi \Rightarrow b_1,\dots,b_{\mathcal{N}'}} 
\langle \Phi | H_0 | \Phi \rangle$,
for the determinants in the subspace. In the same way, given the full $H$,
the method gives a polynomial $\mathcal{E}(\eta,b_1,\dots,b_{\mathcal{N}'}),$
the lowest $\eta$ root of which is the constrained minimum,
$\eta' = \mathrm{Min}_{\Psi \Rightarrow b_1,\dots,b_{\mathcal{N}'}} 
\langle \Psi | H | \Psi \rangle$,
for correlated states in the subspace. Then it is trivial to derive from 
$\mathcal{K}$ and $\mathcal{E}$ a polynomial, 
$\Omega(\omega;b_1,\dots,b_{\mathcal{N}'})$,
for the difference, $\omega=\eta-\kappa$. The diagonalization of
$\mathcal{P} H \mathcal{P}$ then reads,
\begin{equation}
\frac{\partial \kappa}{\partial b_{\beta}} + 
\frac{\partial \omega}{\partial b_{\beta}} = 0,\ \ \beta=1,\dots,\mathcal{N}'.
\label{diago}
\end{equation}
With the ratio, $v_{\beta}=-(\partial \Omega/\partial b_{\beta})/
                            (\partial \Omega/\partial \omega)$, 
representing $\partial \omega/\partial b_{\beta}$, define the
one-body, local potential, 
$v_{\Delta}(r)=\sum_{\beta=1}^{\mathcal{N}'} v_{\beta}\, S_{\beta}(r)$.
Let $\Phi$ be the ground state of
$\mathcal{P} \left[ H_0 + \sum_{i=1}^N v_{\Delta}(r_i) \right] \mathcal{P}.$
Notice that $\langle \Phi | \mathcal{P} S_{\beta} 
\mathcal{P}| \Phi \rangle = \langle \Phi | S_{\beta} | \Phi \rangle.$
Then the energy $E$ of $\Phi$ has derivatives,
\begin{equation}
\partial E/\partial v_{\beta}= 
\int (\Delta \rho+\rho_0)\, S_{\beta} = b_{\beta}+b_{\beta 0},
\label{inhomgnLegndr}
\end{equation}
because of the orthonormality of the $S_{\beta}$'s. The numbers,
$b_{\beta 0}=\int \rho_0\, S_{\beta}$, are easily pretabulated.
The quantities, $v_{\beta}$ and $(b_{\beta}+b_{\beta 0})$, are
Legendre conjugates, and, moreover, 
$\partial/\partial(b_{\beta}+b_{\beta 0})=\partial/\partial b_{\beta}$.
The conditions, Eqs. (\ref{diago}), read as the diagonalization for a
determinant $\Phi$ with the same density $\rho$ as that of the eigenstate
$\Psi$ of $\mathcal{P} H \mathcal{P}$. The potential,
$\mathcal{P} (w_0+v_{\Delta}) \mathcal{P}$, is a KSP valid for the subspace,
up to the convergence of the truncation with $\mathcal{N}'$ terms. 

This polynomial method most often uses a very non local parametrization of 
$\rho,$ that deviates from the quasi-local tradition of the field. In every 
case, our unconventional parametrization of $\rho$  creates a new zoology of 
DFs. Nothing of this zoology is known to us, but its interest is obvious,
since manipulations of polynomials and properties of their roots, including
bounds, are basic subjects. Moreover, extrapolations of polynomials, and 
criticism of such extrapolations, are easy. The number of \textit{available, 
exactly solvable} models is \textit{huge}. It is limited only by computational
power. For nuclei or atoms, the models will be ``radial'' \cite{GirRDF},
somewhat simple. For nuclear physics, our ultimate goal will be to see whether
particle number can be used as a constraint, to generate a mass formula. For
electrons in molecules or extended systems (metals, thin layers, etc.),
however, a necessary algebra of functions of 2 or 3 variables will burden the
models. Anyhow, one can always test whether our polynomials from ``smaller''
models may remain good approximations for ``larger'' ones, if, for instance,
scaling properties can be established. Asymptotic properties of a sequence of
``DF polynomials'' might guide towards derivations of more traditional DFs.
In particular, the polynomial models allow comparisons between the KS and the
true kinetic energies of correlated systems. They also provide explicit terms
for those correlation energies due to interactions.

In conclusion, this algebraic method simplifies density functional theory
into energy minimization under finite numbers of constraints, under very
elementary manipulations of polynomials. It retains all essential information
about the density and all components of the energy. In a forthcoming paper,
we shall investigate a more realistic problem than the toy models used for
this Letter.

SK acknowledges support from the National Research Foundation of South Africa
and thanks CEA/Saclay for hospitality during part of this work.
BG thanks Rhodes University and the University of Johannesburg for their hospitality during part of this work.

\end{document}